# Microsoft Recommenders

## Tools to Accelerate Developing Recommender Systems

Scott Graham
Cloud + AI
Microsoft
Cambridge MA USA
scott.graham@microsoft.com

Jun-Ki Min
Cloud + AI
Microsoft
Cambridge MA USA
jun.min@microsoft.com

Tao Wu
Cloud + AI
Microsoft
Cambridge MA USA
tao.wu@microsoft.com

## ABSTRACT

The purpose of this demonstration is to highlight the content of the Microsoft Recommenders repository and show how it can be used to reduce the time involved in developing recommender systems. The open source repository provides python utilities to simplify common recommender-related data science work as well as example Jupyter notebooks that demonstrate use of the algorithms and tools under various environments.

## CCS CONCEPTS

• Information systems~Recommender systems

## 1 Introduction

The Microsoft Recommenders repository is an open source collection of python utilities and Jupyter notebooks to help accelerate the process of designing, evaluating, and deploying recommender systems. The repository was initially formed by data scientists at Microsoft to consolidate common tools and best practices developed from working on recommender systems in various industry settings. The goal of the tools and notebooks is to show examples of how to effectively build, compare, and then deploy the best recommender solution for a given scenario. Contributions from the community have brought in new algorithm implementations and code examples covering multiple aspects of working with recommendation algorithms. The repository is actively maintained and constantly being improved and extended. It is publicly available on GitHub and licensed through a permissive MIT License to promote widespread use of the tools. The repository maintainers encourage data scientists and developers to contribute their own algorithm implementations, tools, and best practices to continuously improve the content and quality of tools available to the recommender community.

## 2 Repository Components

In this section we describe four main components of the Microsoft Recommenders repository. The first component is the *reco_utils* package, which is sub-divided into separate modules based on different stages of developing recommender systems. While the key capabilities are described below, this is not an exhaustive list. More details on the latest set of functions available and how to use them can be found in the documentation provided with the code. Second is the set of implementations for classical and deep learning-based recommender algorithms. Third, we list the groups of Jupyter notebook examples showing how to use those algorithms. And lastly, we describe the suite of tests used to ensure the utilities and notebooks work as expected.

### 2.1 Recommender Utilities

*2.1.1 Common Utilities.* This submodule contains high-level utilities for defining constants used in most algorithms as well as helper functions for managing aspects of different frameworks: GPU, Spark, and Jupyter notebooks. Helper utilities for common functions dealing with Pandas DataFrames, TensorFlow, and timing algorithm performance are also provided.

*2.1.2 Dataset Utilities.* The dataset submodule includes functions for interacting with Azure Cosmos databases as well as pulling different sizes of sample data to test with, such as the MovieLens [1] and Criteo Display Advertising Challenge [2] datasets. There are methods for splitting data for training, testing, and validation based on random sampling, chronological sampling, or stratified sampling to ensure the same set of users (or items) are in each split. The dataset operations are designed to work with both Pandas and Spark based DataFrames.

*2.1.3 Evaluation Utilities.* The evaluation submodule includes functions for calculating common metrics to evaluate model performance (both in python and Spark). There is a collection of offline test metrics [3] that can be used to evaluate both rating and ranking based approaches.

*2.1.4 Tuning Utilities.* Hyperparameter tuning tools are available to get the best performance out of the algorithms, whether the implementation is on Spark, GPU or CPU. This includes the use of the Neural Network Intelligence toolkit [4], HyperOpt, and Azure HyperDrive service which enable a variety of optimization approaches including: Random Search, Grid Search, Bayesian Optimization and many others.

## 2.2 Recommender Algorithms

The *recommender* module contains implementations of algorithms and integrations with external packages that can be used to evaluate and develop new recommender systems. The following algorithms are currently implemented or integrated using supporting utilities.

**Table 1: List of Implemented Algorithms**

| Algorithm | Type |
| --- | --- |
| Alternating Least Squares | Collaborative Filtering |
| Deep Knowledge Aware Network | Content-based Filtering |
| Extreme Deep Factorization Machine | Hybrid |
| Fast AI Embedding Dot Bias | Collaborative Filtering |
| LightGBM | Content-based Filtering |
| Neural Collaborative Filtering | Collaborative Filtering |
| Reimannian Low-Rank Matrix Completion | Collaborative Filtering |
| Restricted Boltzman Machines | Collaborative Filtering |
| Simple Algorithm for Recommendation | Collaborative Filtering |
| Surprise (Singular Value Decomposition) | Collaborative Filtering |
| Vowpal Wabbit | Hybrid |
| Wide and Deep | Hybrid |

## 2.3 Jupyter Notebook Examples

In the repository *notebooks* folder, there are examples of building recommendation systems written as Jupyter notebooks. The folder structure is broken down to highlight specific aspects of working with recommender systems.

- Quick Start: Simplified workflows for developing recommender models
- Prepare Data: Options for data preparation
- Model: Deep dives on various classical and deep learning recommender algorithms
- Evaluate: Different evaluation approaches for recommender models
- Select and Optimize: How to leverage hyperparameter tuning to improve model performance
- Operationalize: End-to-end examples for putting a recommender model into production and ensuring it satisfies real-world requirements

## 2.4 Testing

Tests are developed for both the utilities and the notebooks to ensure high code quality and functionality for all examples. This project uses unit, smoke, and integration tests with python files and notebooks. Papermill [5] is used extensively when testing the notebooks within the repo. This tool supports adjusting parameters within each notebook then executing and collecting metrics to ensure test cases are passing. Every time a developer makes a pull request to the repository a battery of unit tests is executed to maintain code functionality. Smoke tests and integration tests are run nightly to ensure combinations of components work together and end-to-end examples are functioning.

## 3. Workflow

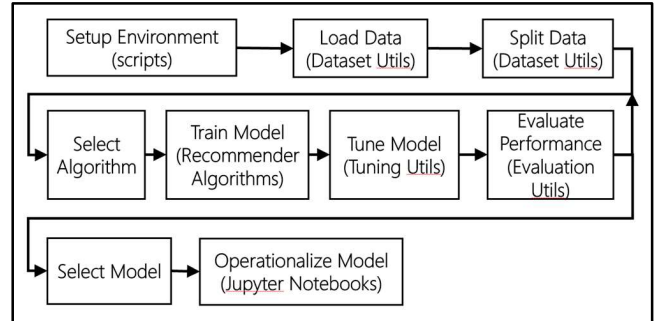

**Figure 1: Example workflow leveraging Recommenders tools**

Figure 1 shows a general workflow for using the tools in Microsoft Recommenders. The first step is to setup the development environment. Several tools are available to automate this process for local development using Anaconda or Docker to manage dependencies for the specific environment desired. Also, scripts are available to simplify setup for using the utilities on Azure Databricks. The Dataset utilities support loading and splitting the data as described earlier. Then the user can select one of the implemented algorithms to train a model. Next the user can tune the hyperparameters used during training and evaluate the performance of the model. Several iterations of algorithm selection and tuning/evaluation can be performed to compare performance across multiple algorithms then select the best approach for the dataset and business needs. The example notebooks can then be adjusted to deploy the selected model for the desired end production environment.

## ACKNOWLEDGMENTS
This work is supported by Microsoft Cloud + AI group in collaboration with both internal and external contributors [6]. Special thanks to Andreas Argyriou, Miguel González-Fierro, and Le Zhang for their work in the development of the repository.